# Ultra-stable optical clock with two cold-atom ensembles

M. Schioppo[1,2,3], R. C. Brown[1], W. F. McGrew[1,2], N. Hinkley[1,2], R. J. Fasano[1,2],

K. Beloy[1], T. H. Yoon[1,4], G. Milani[1,5,6], D. Nicolodi[1], J. A. Sherman[1], N. B. Phillips[1],

C. W. Oates[1] and A. D. Ludlow[1*]

[1]National Institute of Standards and Technology, 325 Broadway, Boulder, Colorado 80305, USA

[2]Department of Physics, University of Colorado, Boulder, CO 80309, USA.

[3]Institute für Experimentalphysik, Heinrich-Heine-Universität Düsseldorf, Germany

[4]Department of Physics, Korea University, 145 Anam-ro, Seongbuk-gu, Seoul 02841, Korea

[5]Instituto Nazionale di Ricerca Metrologica, Strada delle Cacce 91, 10135 Torino, Italy

[6]Politecnico di Torino, Corso duca degli Abruzzi 24, 10125 Torino, Italy

**Atomic clocks based on optical transitions are the most stable, and therefore precise, timekeepers available. These clocks operate by alternating intervals of atomic interrogation with 'dead' time required for quantum state preparation and readout. This non-continuous interrogation of the atom system results in the Dick effect, an aliasing of frequency noise of the laser interrogating the atomic transition[1,2]. Despite recent advances in optical clock stability achieved by improving laser coherence, the Dick effect has continually limited optical clock performance. Here we implement a robust solution to overcome this limitation: a zero-dead-time optical clock based on the interleaved interrogation of two cold-atom ensembles[3]. This clock exhibits vanishingly small Dick noise, thereby achieving an unprecedented fractional frequency instability of $6 \times 10^{-17}/\sqrt{\tau}$ for an averaging time $\tau$ in seconds. We also consider alternate dual-atom-ensemble schemes to extend laser coherence and reduce**

**the standard quantum limit of clock stability, achieving a spectroscopy line quality factor $Q > 4 \times 10^{15}$.**

An optical atomic clock operates by tuning the frequency of a laser (optical local oscillator, OLO) into resonance with a narrowband, electronic transition in an atomic system[4,5]. By so doing, the intrinsic 'ticking' rate of the atom (~$10^{15}$ Hz), given by the phase evolution of the electronic wave function, is transferred to the laser field for use as a timebase. Any noise influencing this transfer process compromises the resulting laser frequency stability, and thus the timekeeping precision. Optical clock stability is typically compromised by two classes of noise: atomic detection and OLO-induced noise.

Atomic detection noise encapsulates processes associated with atomic state measurement. The ultimate limit arises from quantum projection noise (QPN)[6], as an atom's state collapses non-deterministically to a particular eigenstate. Photon shot noise associated with the measurement technique (e.g. laser-induced fluorescence) also contributes, as does technical noise derived from fluctuations in: (i) the optical field used to measure the atomic state, (ii) the number of atoms participating, (iii) the opto-electronic signal ultimately being measured. Adding these noise terms, a simplified expression for the fractional clock instability resulting from atomic detection can be formed[4]:

$$\sigma_{\text{atom}}(\tau) \approx \frac{1}{\pi Q} \sqrt{\frac{T_\text{C}}{\tau}} \sqrt{\frac{1}{N} + \frac{1}{Nn} + \delta_{\text{det}}^2} \tag{1}$$

where $Q$ is the line quality factor (ratio of electronic transition frequency to linewidth), $T_\text{C}$ is the clock cycle time, $N$ is the number of atoms being interrogated with $n$ photons detected per atom, and $\delta_{\text{det}}$ characterizes technical detection fluctuations manifest as a fractional variation in the normalized atomic excitation. For the optical lattice clocks considered here, typical conditions can be $T_C \approx 0.5$ s, $Q \approx 10^{14}$, $N \approx 10^4$, and $n \gg 1$. Under these conditions, the fundamental limit to atomic detection noise given by QPN (first term in Eq. 1) yields a potential clock instability of $2 \times 10^{-17}/\sqrt{\tau}$. Such performance makes the optical lattice clock a powerful tool for a variety of precision tests of fundamental constants[4], general relativity[7], and even searches for dark matter[8,9].

Regrettably, atomic detection noise is typically exceeded by noise originating from the OLO. During quantum state preparation and readout, OLO frequency fluctuations are not observed by the atomic system. Consequently, fluctuations at harmonic frequencies of $1/T_C$ are aliased by this periodic atomic interrogation. This aliased noise, i.e. Dick noise[1], is then improperly compensated in the laser tuning process, compromising the optical clock stability. The Dick instability, for a clock duty cycle near 50%, can be approximated as[4]:

$$\sigma_{\text{Dick}}(\tau) \approx \frac{\sigma_{\text{OLO}}}{\sqrt{2\ln 2}} \sqrt{\frac{T_C}{\tau}} \left| \frac{\sin(\pi T_F/T_C)}{\pi T_F/T_C} \right|$$

(2)

where $T_F$ is the Ramsey free-evolution-time and $\sigma_{\text{OLO}}$ is the OLO flicker frequency instability. The OLO usually comprises a laser pre-stabilized to an isolated, high-finesse Fabry-Perot cavity. For typical operating conditions and a state-of-the-art thermal-noise-limited cavity, Dick noise remains twice as large as the QPN limit. Consequently, the optical clock stability is mostly set by the OLO rather than the atomic system. Advances in the stability of optical clocks have thus been realized using iteratively higher performing OLOs[10-19]. In some cases, the OLO becomes experimentally more complex than the atomic system it interrogates, challenging future applications of optical clocks outside the laboratory[20], including relativistic geodesy[21,22] and space-borne tests of general relativity[23].

Here we implement two optical clocks using ytterbium atoms trapped in an optical lattice interrogated by a shared OLO (Fig. 1 and Methods). Table 1 shows the instability budget of these clocks. With $N \approx 5,000-10,000$ atoms, the single clock QPN is $1 \times 10^{-17}/\sqrt{\tau}$, while technical fluctuations in the atomic detection measured to be $\delta_{det} \approx 3\%$ contribute an instability of $3 \times 10^{-17}/\sqrt{\tau}$. Additional noise $\sigma_{phase}$ arising from length fluctuations in the optical path between the OLO and the two atomic systems is also considered.

To highlight the importance of Dick noise, we first perform atomic interrogation in an anti-synchronized fashion so that the Ramsey free-evolution-time of one system ($T_F = 240$ ms) overlaps the dead-time of the other system (Fig. 2a). This ensures maximum sensitivity to the OLO aliasing process. Given the measured flicker frequency instability of our OLO $\leq 1.5 \times 10^{-16}$, the Dick effect for one system is calculated to be $7 \times 10^{-17}/\sqrt{\tau}$. As displayed in Fig. 2a, in this anti-synchronized configuration we measure a single-clock instability of $1.4 \times 10^{-16}/\sqrt{\tau}$ in good agreement with Table 1. This demonstrates that use of a state-of-the-art OLO affords unprecedented optical clock stability[10-15]. Nevertheless, Dick noise continues to dominate the achieved level of performance. Improvements could be made by enhancing OLO coherence, which also affords longer spectroscopy times. An alternative approach reduces dead-time with multiple non-destructive measurements repeated after a single period of atomic preparation[24]. However, Fig. 3 highlights that in both these cases and for the conditions considered here, performance is still limited above the QPN limit.

A distinct optical clock architecture combining two atomic ensembles has the potential to nearly eliminate aliased OLO noise[3,25]. Using the same anti-synchronized interrogation, measurements of the atomic response of each system are now combined to derive a single, shared error signal and frequency correction to the OLO. The combined atomic system provides quasi-continuous monitoring of the OLO frequency, which significantly reduces the aliasing effect. This can be described quantitatively with the sensitivity function (Fig. 2b), defining the atomic response to OLO frequency fluctuations over time. Each atomic ensemble exhibits unit sensitivity during its Ramsey free-evolution-time and zero sensitivity during its dead-time. When merged together to yield a two-ensemble optical clock, the sensitivity remains constant at all times except for small variations during the short Ramsey pulses, and thus provides significantly reduced susceptibility to OLO frequency noise at harmonics of $1/T_C$. Figure 2b shows the shared-frequency-correction used in this zero-dead-time (ZDT) stabilization process. For the same experimental conditions listed above, the Dick instability is computed to be $< 3 \times 10^{-18}/\sqrt{\tau}$, now significantly below QPN. Further reduction could be realized by fine adjustment of the relative timing and shape of the Ramsey pulses, to achieve an even more uniform composite-sensitivity-function[3].

A direct measurement of the ZDT clock instability would be possible by comparing two independent ZDT systems[26]. Lacking the four atomic ensembles needed for such a measurement, we evaluate the clock stability with an alternative scheme using synchronized interrogation of two atomic systems, as shown in Fig. 2c, using a shared OLO[16]. Due to the dead-time and the corresponding time-dependent sensitivity function of each atomic system, both individually suffer from Dick noise. However, because interrogation is performed with a single OLO in a synchronized fashion, the Dick noise is correlated and common-mode rejected in the comparison between the two clocks. As a result, the synchronized measurement and ZDT clock are sensitive to the same noise processes, as displayed in Table 1, and thus the synchronized measurement can be used to experimentally evaluate the ZDT clock. The resulting synchronized measurement stability is shown in Fig. 2c, averaging down as $8\times10^{-17}/\sqrt{\tau}$. For a white frequency noise process, the ZDT clock is $\sqrt{2}$ more stable than the synchronized measurement because it accumulates atomic measurements and OLO corrections twice as fast, yielding a ZDT instability of $6\times10^{-17}/\sqrt{\tau}$. With the Dick noise suppressed, it is now possible to reach an instability level of $1\times10^{-18}$ in a mere few thousand seconds, an order of magnitude faster than the previous clock stability demonstrations[10-15].

Beyond reducing Dick noise, two atomic systems can also be combined in alternate configurations to decrease QPN. Recent proposals use one system in a succession of short interrogation periods to prestabilize the OLO such that a second system can then be interrogated over a longer period than the OLO coherence would otherwise permit[27,28]. Fig. 4 displays the experimental realization toward one such scheme with the OLO pre-stabilized on one atomic system to increase the interrogation time up to 2.4 s and 4 s on the second atomic system. This leads to Fourier limited Ramsey fringes with linewidths of $(210\pm20)$ mHz and $(120\pm20)$ mHz respectively, with a constrast of 40%, realizing a $Q>4\times10^{15}$, the highest in an optical lattice clock system. The longer interrogation time and higher $Q$ thus affords reduced QPN (see Eq. 1). Because this approach fails to eliminate dead time, its inherent Dick noise suppression is modest compared to that of the ZDT configuration. Therefore, either the ZDT configuration or the long interrogation scheme could be optimal, depending on the relative magnitude of Dick noise and QPN. Alternatively, if atomic phase correlations/entanglement could be propagated between two atomic ensembles in a ZDT configuration, the OLO phase could be continuously tracked, fundamentally improving the noise averaging process and thus the clock stability.

We have demonstrated a robust, universal scheme for eliminating Dick noise in an optical clock. This is possible by utilizing the atomic system rather than relying on the OLO coherence. By so doing, we have observed unprecedented clock stability, enabling very high precision with averaging times much shorter than previously possible. With OLO-induced noise removed, optical lattice clocks can reach performance given by atomic detection noise and its fundamental QPN limit. Attention can thus turn to reducing QPN, as well as spin-squeezing techniques that enable performance beyond QPN[29,30]. The improved stability and reduced measurement time demonstrated here greatly enhances atomic clock utility in tests of fundamental physics, opening the door to $10^{-19}$-level timekeeping precision. Furthermore, because of the relaxed requirements on the OLO performance, this scheme can reduce the total complexity of the optical clock, extending its usability outside the laboratory.

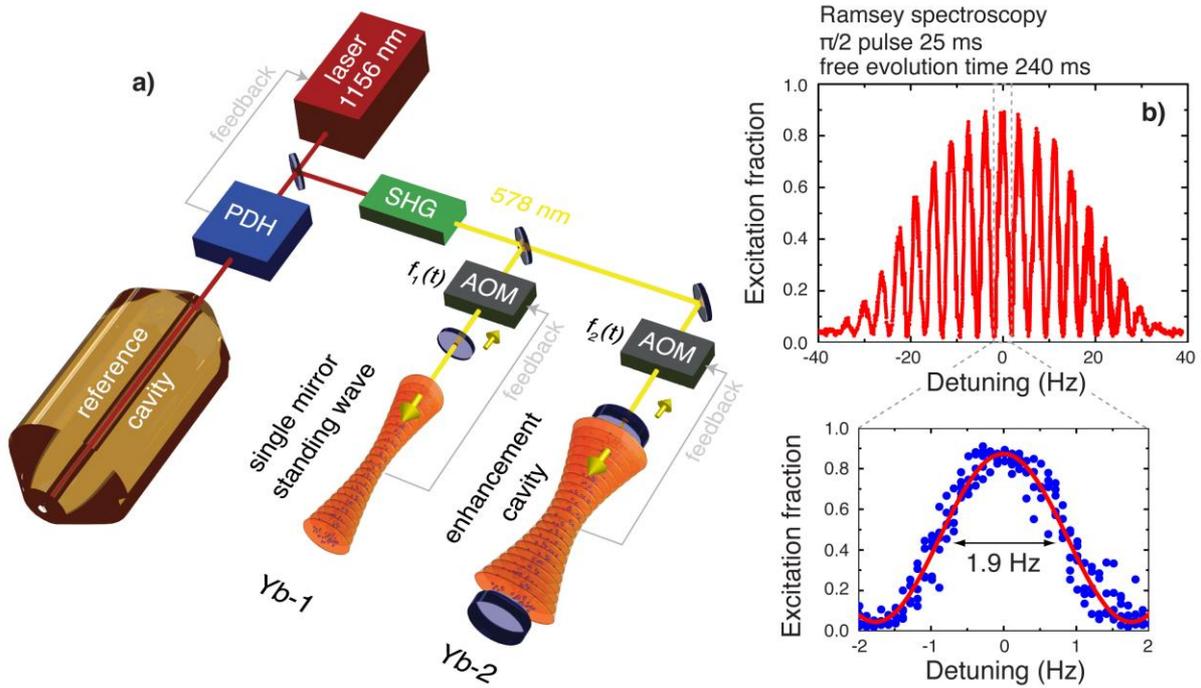

**Figure 1 | Experimental scheme. a,** Laser light generated by a quantum dot laser at 1156 nm is Pound-Drever-Hall (PDH) stabilized to a high-finesse optical cavity, frequency doubled through a Second-Harmonic-Generation (SHG) crystal waveguide to reach the atomic clock transition at 578 nm, split between two atomic systems (Yb-1 and Yb-2) and tuned in resonance by two independent frequency shifters (AOMs). At the lattice, a fraction of the light at 578 nm is reflected back to interferometrically detect and remove environmental phase noise accumulated in the optical path from the cavity to the lattice. **b**, Ramsey spectrum under typical conditions used to frequency stabilize the clock laser on the atomic transition. Each fringe has a linewidth of 1.9 Hz, with the central component having a contrast >90%. The shown spectra have no averaging.

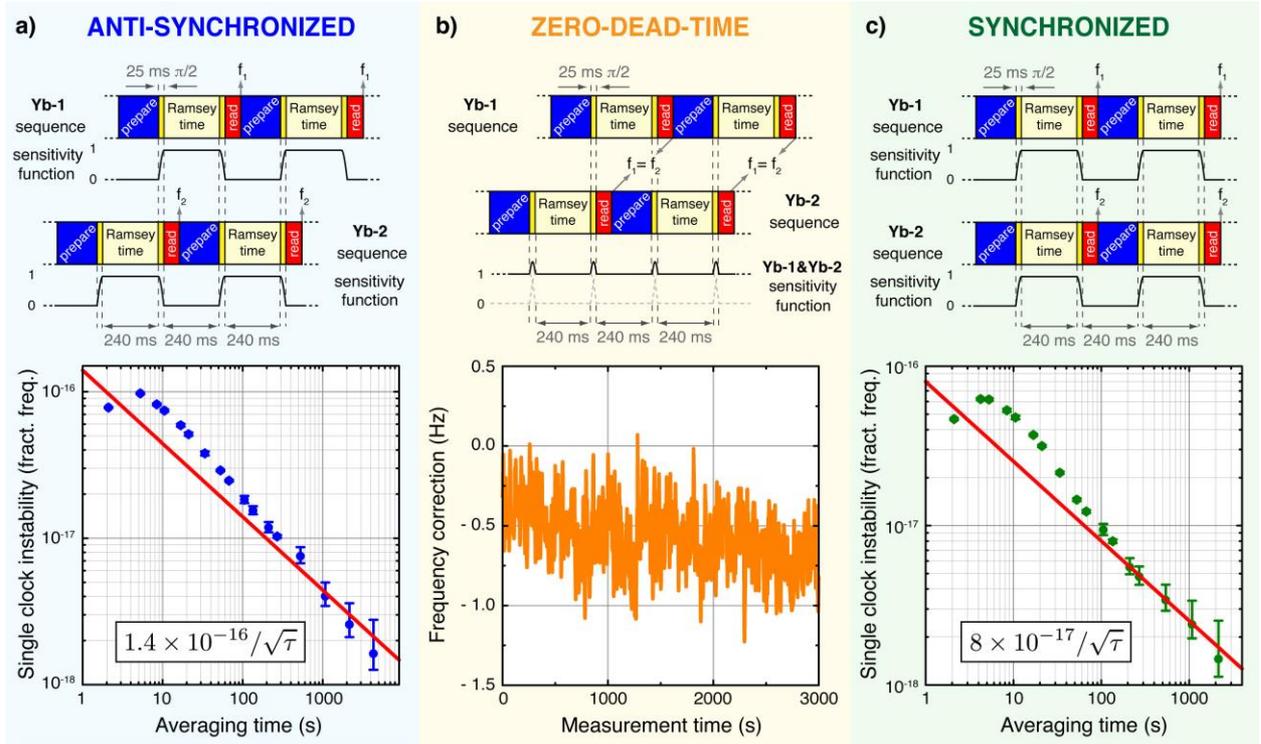

**Figure 2 | Timing sequence, sensitivity function and instability measurement for the three relevant schemes studied here. a,** The atomic systems are interrogated anti-synchronously and the corrections applied independently to their respective AOMs. The Allan deviation of the AOMs frequency difference $f_1(t)-f_2(t)$ results in a measured single clock instability of $1.4 \times 10^{-16}/\sqrt{\tau}$. **b,** In the zero-dead-time (ZDT) configuration the correction derived from the interrogation of one system is applied phase-continuously during the interrogation of the second system such that the clock laser continuously interrogates the atoms and is corrected every half of the cycle time, eliminating the source of the Dick effect. Yb-1 and Yb-2 form a composite system, with a shared correction frequency (stabilizing the free-running clock laser to the composite system) of $f_1(t)=f_2(t)$. **c,** Synchronized spectroscopy offers a method to evaluate the instability of the ZDT operation since the OLO noise is common-mode-rejected, virtually eliminating the Dick effect. The resulting synchronized measurement stability is $8 \times 10^{-17}/\sqrt{\tau}$. In the ZDT scheme, the observed stability will be improved by $\sqrt{2}$ because the atomic measurements and OLO corrections accumulate twice as often, leading to an estimated ZDT clock instability of $6 \times 10^{-17}/\sqrt{\tau}$.

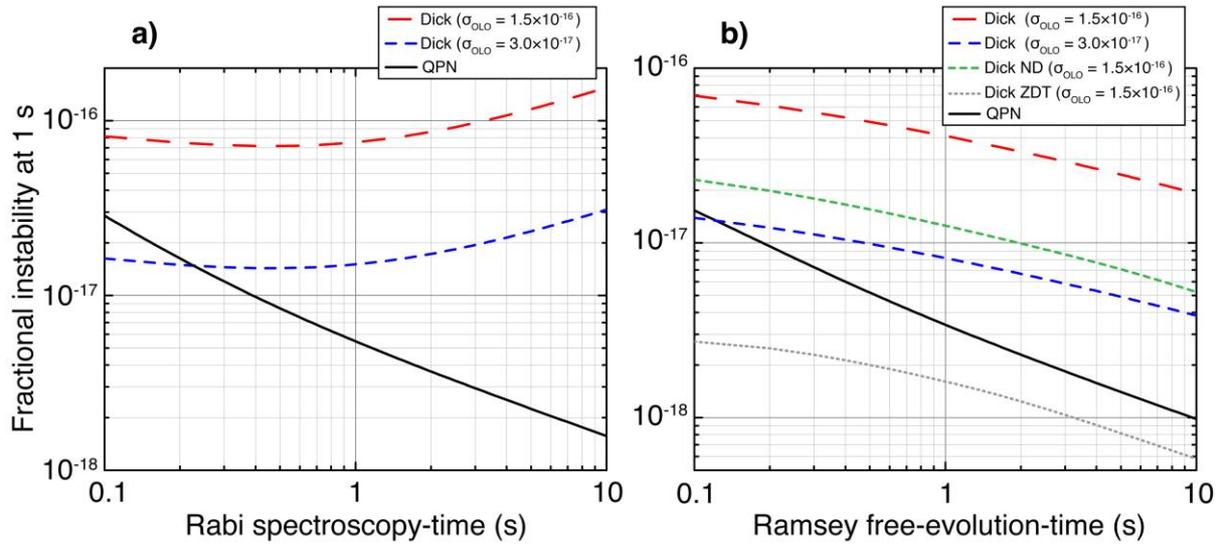

**Figure 3 | Computed Dick and QPN instability at 1 s assuming an atom number of 10,000. a,** The contribution of the Dick effect is evaluated as a function of Rabi spectroscopy-time (dead-time fixed to 240 ms) for the measured cavity flicker frequency instability of $1.5 \times 10^{-16}$ (red long-dashed curve) and for a prospective improved cavity with flicker frequency instability of $3.0 \times 10^{-17}$ (blue medium-dashed curve). Note that for Rabi spectroscopy, as the OLO flicker frequency noise is aliased, the Dick limit can get worse at longer spectroscopy time. For comparison the relative QPN limit is displayed for the Rabi scheme (black solid curve). **b,** Here the Dick instability is evaluated as a function of the Ramsey free-evolution-time for the case of cavity flicker frequency instability of $1.5 \times 10^{-16}$ (red long-dashed curve) and $3.0 \times 10^{-17}$ (blue medium-dashed curve) both with a fixed dead-time of 240 ms. The Dick effect is calculated for non-destructive Ramsey spectroscopy (green short-dashed curve) assuming 100 consecutive measurements with 40 ms read time, cavity flicker frequency instability of $1.5 \times 10^{-16}$ and a fixed initial dead-time of 240 ms. Dick instability is also evaluated for Ramsey zero-dead-time (gray dotted curve) for a fixed duty-cycle of 50%. The QPN limit for Ramsey spectroscopy is displayed for comparison (black solid curve). The duration of the $\pi/2$ pulses is fixed to 25 ms.

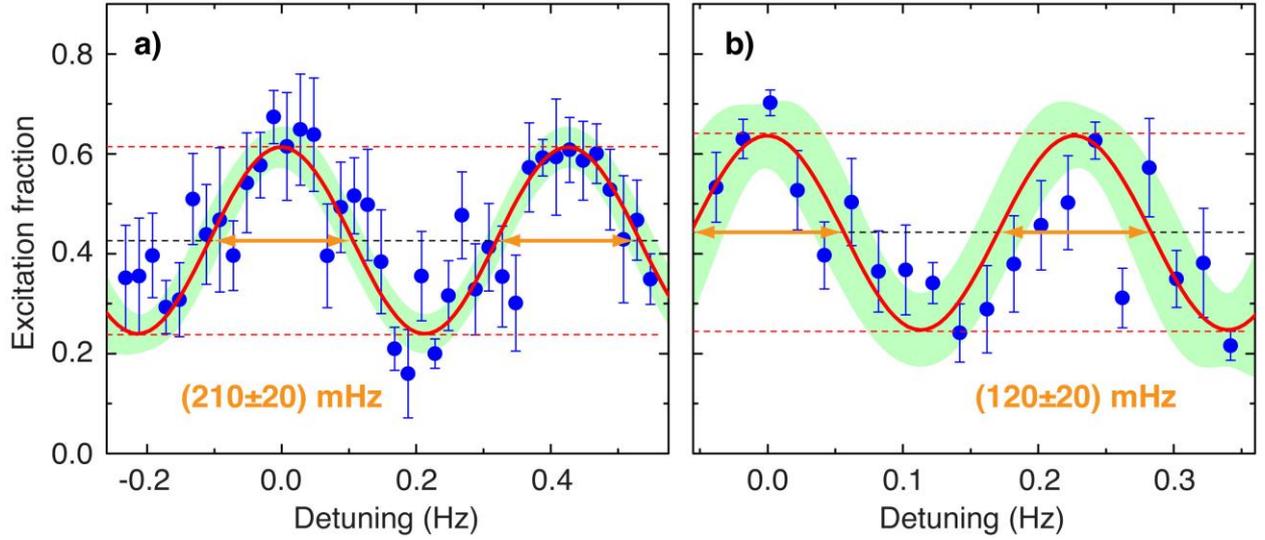

**Figure 4 | Long interrogation Ramsey spectroscopy with the OLO pre-stabilized by one atomic system with cycle and free-evolution time of 240 ms and 110 ms respectively. To demonstrate the repeatability of these narrow line spectra, here we show the average of seven consecutive laser detuning sweeps. The measured atomic excitation versus detuning is shown by blue circles. A fit to the data is given by the red solid line, with the 95% confidence band indicated by the green shadow. a,** Ramsey spectroscopy with the pre-stabilized OLO for a free-evolution-time on the second atomic system of 2.4 s, leading to a (Fourier-limited) fringe linewidth of $(210 \pm 20)$ mHz with 40% contrast. **b.** Here the free evolution time for Ramsey spectroscopy on the second system is 4 s, corresponding to a (Fourier-limited) fringe linewidth of $(120 \pm 20)$ mHz with 40% contrast.

**Table 1 | Single clock instability budget for different operational configurations, in units of $10^{-17}/\sqrt{\tau}$. Individual contributions from quantum-projection noise, technical noise in the detection, residual noise from phase-stabilization of the clock laser and the Dick effect are included, respectively. Total gives the quadrature sum of the different noise terms (see Methods).**

|  | Anti-synchronized | Synchronized | Zero-dead-time |
|---|---|---|---|
| $\sigma_{\text{QPN}}$ | 1 | 1 | 1/√2 |
| $\sigma_{\text{technical}}$ | 3 | 3 | 3/√2 |
| $\sigma_{\text{phase}}$ | 2 | 2 | 2 |
| $\sigma_{\text{Dick}}$ | 7×√2 | common-mode rejected | < 0.3 |
| $\sigma_{\text{Total}}$ | **11** | **4** | **3** |

**Methods**

**Optical local oscillator (OLO).** The clock laser is based on a quantum dot laser at 1156 nm amplified and frequency stabilized on a high-finesse Fabry-Perot cavity with a length of 29 cm. The power storage time of the reference cavity is 270 μs, corresponding to a finesse of 877 000. The radius of curvature of each mirror is 10.5 m to average the Brownian noise of the dielectric mirrors on a large area. The $1/e^2$ intensity beam diameter on the mirrors is 1.34 mm. The frequency instability of the clock laser has been measured using the atomic transition as an independent frequency discriminator at the fractional level of $\leq 1.5 \times 10^{-16}$ in the averaging time interval 1 s to 1000 s. This measurement provides an upper limit since it also includes atomic detection noise. The high finesse allows to passively reduce the technical instability due to residual amplitude modulation of the phase modulator to $\leq 2 \times 10^{-17}$ in the averaging time interval 1 s to 100 s. The reference cavity is actively temperature stabilized at the null-coefficient-of-thermal-expansion temperature of 35 °C. Three layers of nested in-vacuum thermal radiation shields provide passive thermal insulation with an estimated characteristic time response of a few days. The drift of the cavity has been measured using the atomic transition as a reference on a time scale of one year. During this interval a constant drift of -35 mHz/s has been observed and linearly compensated with residual non-linear deviations not exceeding 1 mHz/s in one day. The sign and the value of the drift is compatible with the creeping of the ultra-low-expansion glass spacer. The sub-harmonic light is frequency doubled to 578 nm through a single pass nonlinear wave guide, providing 5 mW power available for the atomic system.

**Ytterbium optical lattice.** Both atomic systems Yb-1 and Yb-2 trap ultracold ytterbium in a one-dimensional optical lattice operating at the 'magic' wavelength, where Stark shifts from the lattice confinement are matched for both atomic states used for the frequency standard ($^1S_0$ and $^3P_0$)[31,32]. The lattice laser frequency is held to the 'magic' value by stabilization to a reference Fabry-Perot cavity and by occasional measurement of its absolute frequency with an optical frequency comb. The lattice standing wave in Yb-1 is generated by simple retroreflection of an incident laser, while that for Yb-2 utilizes an enhancement cavity to generate optical power buildup of more than one hundred times the incident laser power and thus enabling use of a large cavity mode diameter to increase the trapping volume. While the trap depths utilized in each lattice can vary depending on need, for measurements highlighted here, trap depths of 15 $\mu$K are employed, and with atomic temperatures of 5-10 $\mu$K. The number of atoms trapped, influencing the relevant quantum projection noise, is 5,000 for Yb-1 and 10,000 for Yb-2. Spectroscopy on the $^1S_0$-$^3P_0$ transition is carried out in both the Lamb-Dicke and well-resolved sideband regimes, eliminating residual Doppler and recoil effects[33]. After spectroscopy, atomic population in both $^1S_0$ ground and $^3P_0$ excited electronic states is measured using laser fluorescence shelving detection[34-36]. The excitation signal, normalized against atom number, is used to stabilize the OLO frequency onto resonance with the lattice-trapped atoms. The choice of Ramsey spectroscopy maximizes the line $Q$ factor for a given spectroscopy time of about one half cycle. Indeed the Fourier-limited linewidth for Ramsey spectroscopy is about 1.6 times narrower than that for Rabi spectroscopy.

**Rejection of aliased noise.** In the time domain, each half of the interleaved, zero-dead-time clock remains sensitive to the Dick effect. However, because the same OLO pulse serves as the first Ramsey pulse in one system and the second Ramsey pulse in the other, the effect of the aliased noise in the two systems has the same magnitude but opposite sign[37]. Thus when added together, the Dick noise of the two systems effectively cancel. In order for this suppression to work effectively, OLO noise must be correlated between the two atomic ensembles. For this reason, phase noise originating from optical path fluctuations between the OLO and the atomic reference frame, given by the standing wave of each optical lattice, must be actively compensated with an acousto-optic modulator[38,39]. Using a Mach-Zehnder interferometer, we have measured this compensation to be effective at $\leq 2\times 10^{-17}$ instability in 1 s, which can be further improved. In order to ensure that no $\pi$-radian optical phase shifts occur between Ramsey spectroscopy pulses (due to the $2\pi$ ambiguity of the round-trip phase detection process) some OLO light remains on during the Ramsey free-evolution-time. This light is far-detuned from the atomic transition to prevent excitation, and enables the active phase stabilizer to maintain phase continuity between Ramsey pulses.

For Dick suppression to work well, the Ramsey pulses delivered to the two atomic systems must be matched in timing and shape. The Ramsey pulse overlap is realized at the μs level. Furthermore, there is no discernible difference between the square OLO pulses to the two systems, at the 0.1% level. Finally, because each atomic interrogation remains sensitive to a relatively large first-order Zeeman shift, differential magnetic field fluctuations between the two atomic systems must be maintained below the mG level. These requirements are important for both the ZDT clock, as well as the synchronized measurements between two atomic systems used to assess ZDT performance.

In order to amplify the observed Dick suppression achieved in the synchronized measurements and possibly in the ZDT clock, we intentionally added white frequency noise onto the OLO while leaving the other operational parameters of the clock unchanged. The noise amplitude was sufficient to increase the anti-synchronized measurement stability to $7 \times 10^{-16}/\sqrt{\tau}$, while the synchronized measurement remained below $1.5 \times 10^{-16}/\sqrt{\tau}$.

**Dick noise in the anti-synchronized measurement.** We access the single clock instability in the anti-synchronized measurement by dividing the instability of the measured frequency difference by $\sqrt{2}$, with each system contributing equal and uncorrelated noise. However, in the anti-synchronized configuration, Dick noise between the two systems is correlated, requiring a linear sum for this noise term. This effect is reflected in the instability budget (Table 1) by multiplying the contribution of Dick noise by $\sqrt{2}$.

**Acknowledgements**

The authors acknowledge DARPA, QuASAR, NASA Fundamental Physics, and NIST for financial support. R.C.B. acknowledges support from the National Research Council Research Associateship program. We also thank T. Fortier, F. Quinlan, and S. Diddams for femtosecond optical frequency comb measurements.

**Author contributions**

M.S., R.C.B., W.F.M., R.J.F., G.M., D.N. and A.D.L. carried out the instability measurements reported here. M.S. and A.D.L. constructed the clock laser. W.F.M., T.H.Y and A.D.L. contributed to the optimization of clock laser performance. J.A.S. constructed the DDS system for precise cavity drift compensation. R.C.B., N.H., T.H.Y., W.F.M., R.J.F., G.M. and A.D.L. were responsible for the operation of Yb-1 and Yb-2 systems and its phase noise cancellation. K.B. contributed to the evaluation of the instability budget. C.W.O. and A.D.L. supervised this work. All authors contributed to the final manuscript.

**Additional information**

The authors declare no competing financial interests. Reprints and permission information is available online at http://npg.nature.com/reprintsandpermissions/. Correspondence and requests for materials should be addressed to A.D.L. This is the work of U.S. government and not subject to copyright.

**Supplementary information**

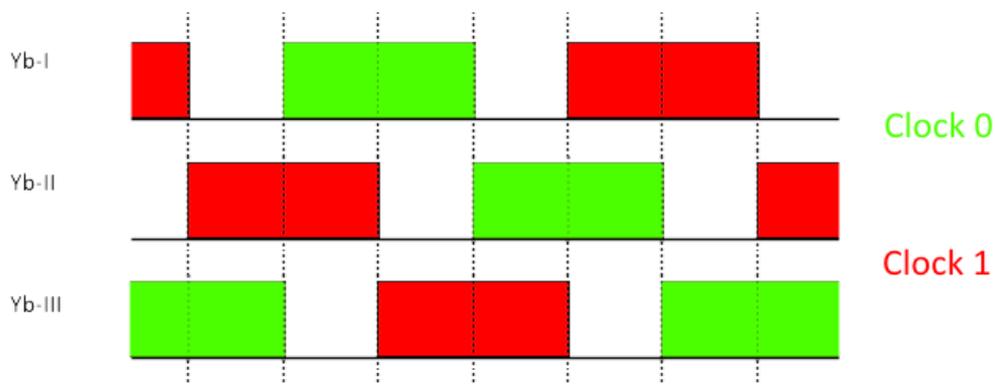

**Figure 5 | Two independent zero-dead-time clocks utilizing three atomic systems.**

As noted in the main text, a direct method to evaluate the ZDT clock instability is by comparison of 2 ZDT clocks, requiring four atomic systems with 50% duty cycle. However, we note that it is also possible to make $m$ zero-dead-time error signals-"clocks"-out of $n$ atomic systems, given a duty cycle $m/n$. Figure 5 demonstrates an operational sequence in which two zero-dead-time clocks are stabilized to three atomic systems. A duty cycle of 2/3 is achievable with our current OLO and atomic preparation sequence. While such a technique can be useful for evaluation of short and medium timescale stability performance of the ZDT clock, long timescale drifts on the atom frequency would be common-mode between the ZDT systems, preventing their accurate determination.